\documentclass{aastex61}

\received{July 26, 2017}
\revised{October 18, 2017}
\accepted{\today}
\submitjournal{ApJ}

\shorttitle{\ IceCube coincidence search in Super-Kamiokande}
\shortauthors{Abe et al.}

\begin{document}

\title{Search for an excess of events in the Super-Kamiokande detector
  in the directions of the astrophysical neutrinos reported
  by the IceCube Collaboration}

\correspondingauthor{Erin O'Sullivan}
\email{erin.osullivan@fysik.su.se}

\newcommand{\AFFicrr}{\affiliation{Kamioka Observatory, Institute for Cosmic Ray Research, University of Tokyo, Kamioka, Gifu 506-1205, Japan}}
\newcommand{\AFFkashiwa}{\affiliation{Research Center for Cosmic Neutrinos, Institute for Cosmic Ray Research, University of Tokyo, Kashiwa, Chiba 277-8582, Japan}}
\newcommand{\AFFipmu}{\affiliation{Kavli Institute for the Physics and
Mathematics of the Universe (WPI), The University of Tokyo Institutes for Advanced Study,
University of Tokyo, Kashiwa, Chiba 277-8583, Japan }}
\newcommand{\AFFmad}{\affiliation{Department of Theoretical Physics, University Autonoma Madrid, 28049 Madrid, Spain}}
\newcommand{\AFFubc}{\affiliation{Department of Physics and Astronomy, University of British Columbia, Vancouver, BC, V6T1Z4, Canada}}
\newcommand{\AFFbu}{\affiliation{Department of Physics, Boston University, Boston, MA 02215, USA}}
\newcommand{\AFFbnl}{\affiliation{Physics Department, Brookhaven National Laboratory, Upton, NY 11973, USA}}
\newcommand{\AFFuci}{\affiliation{Department of Physics and Astronomy, University of California, Irvine, Irvine, CA 92697-4575, USA }}
\newcommand{\AFFcsu}{\affiliation{Department of Physics, California State University, Dominguez Hills, Carson, CA 90747, USA}}
\newcommand{\AFFcnm}{\affiliation{Department of Physics, Chonnam National University, Kwangju 500-757, Korea}}
\newcommand{\AFFduke}{\affiliation{Department of Physics, Duke University, Durham NC 27708, USA}}
\newcommand{\AFFfukuoka}{\affiliation{Junior College, Fukuoka Institute of Technology, Fukuoka, Fukuoka 811-0295, Japan}}
\newcommand{\AFFgifu}{\affiliation{Department of Physics, Gifu University, Gifu, Gifu 501-1193, Japan}}
\newcommand{\AFFgist}{\affiliation{GIST College, Gwangju Institute of Science and Technology, Gwangju 500-712, Korea}}
\newcommand{\AFFuh}{\affiliation{Department of Physics and Astronomy, University of Hawaii, Honolulu, HI 96822, USA}}
\newcommand{\AFFicl}{\affiliation{Department of Physics, Imperial College London , London, SW7 2AZ, United Kingdom }}
\newcommand{\AFFkek}{\affiliation{High Energy Accelerator Research Organization (KEK), Tsukuba, Ibaraki 305-0801, Japan }}
\newcommand{\AFFkobe}{\affiliation{Department of Physics, Kobe University, Kobe, Hyogo 657-8501, Japan}}
\newcommand{\AFFkyoto}{\affiliation{Department of Physics, Kyoto University, Kyoto, Kyoto 606-8502, Japan}}
\newcommand{\AFFliv}{\affiliation{Department of Physics, University of Liverpool, Liverpool, L69 7ZE, United Kingdom}}
\newcommand{\AFFmiyagi}{\affiliation{Department of Physics, Miyagi University of Education, Sendai, Miyagi 980-0845, Japan}}
\newcommand{\AFFnagoya}{\affiliation{Institute for Space-Earth Enviromental Research, Nagoya University, Nagoya, Aichi 464-8602, Japan}}
\newcommand{\AFFkmi}{\affiliation{Kobayashi-Maskawa Institute for the Origin of Particles and the Universe, Nagoya University, Nagoya, Aichi 464-8602, Japan}}
\newcommand{\AFFpol}{\affiliation{National Centre For Nuclear Research, 00-681 Warsaw, Poland}}
\newcommand{\AFFsuny}{\affiliation{Department of Physics and Astronomy, State University of New York at Stony Brook, NY 11794-3800, USA}}
\newcommand{\AFFokayama}{\affiliation{Department of Physics, Okayama University, Okayama, Okayama 700-8530, Japan }}
\newcommand{\AFFosaka}{\affiliation{Department of Physics, Osaka University, Toyonaka, Osaka 560-0043, Japan}}
\newcommand{\AFFox}{\affiliation{Department of Physics, Oxford University, Oxford, OX1 3PU, United Kingdom}}
\newcommand{\AFFqmul}{\affiliation{School of Physics and Astronomy, Queen Mary University of London, London, E1 4NS, United Kingdom}}
\newcommand{\AFFregina}{\affiliation{Department of Physics, University of Regina, 3737 Wascana Parkway, Regina, SK, S4SOA2, Canada}}
\newcommand{\AFFseoul}{\affiliation{Department of Physics, Seoul National University, Seoul 151-742, Korea}}
\newcommand{\AFFsheff}{\affiliation{Department of Physics and Astronomy, University of Sheffield, S10 2TN, Sheffield, United Kingdom}}
\newcommand{\AFFshizuokasc}{\affiliation{Department of Informatics in
Social Welfare, Shizuoka University of Welfare, Yaizu, Shizuoka, 425-8611, Japan}}
\newcommand{\AFFstfc}{\affiliation{STFC, Rutherford Appleton Laboratory, Harwell Oxford, and Daresbury Laboratory, Warrington, OX11 0QX, United Kingdom}}
\newcommand{\AFFskk}{\affiliation{Department of Physics, Sungkyunkwan University, Suwon 440-746, Korea}}
\newcommand{\AFFtokyo}{\affiliation{The University of Tokyo, Bunkyo, Tokyo 113-0033, Japan }}
\newcommand{\AFFtodai}{\affiliation{Department of Physics, University of Tokyo, Bunkyo, Tokyo 113-0033, Japan }}
\newcommand{\AFFtit}{\affiliation{Department of Physics,Tokyo
Institute of Technology, Meguro, Tokyo 152-8551, Japan }}
\newcommand{\AFFtus}{\affiliation{Department of Physics, Faculty of Science and Technology, Tokyo University of Science, Noda, Chiba 278-8510, Japan }}
\newcommand{\AFFtoronto}{\affiliation{Department of Physics, University of Toronto, ON, M5S 1A7, Canada }}
\newcommand{\AFFtriumf}{\affiliation{TRIUMF, 4004 Wesbrook Mall, Vancouver, BC, V6T2A3, Canada }}
\newcommand{\AFFtokai}{\affiliation{Department of Physics, Tokai University, Hiratsuka, Kanagawa 259-1292, Japan}}
\newcommand{\AFFtsinghua}{\affiliation{Department of Engineering Physics, Tsinghua University, Beijing, 100084, China}}
\newcommand{\AFFynu}{\affiliation{Faculty of Engineering, Yokohama National University, Yokohama, 240-8501, Japan}}
\newcommand{\AFFuw}{\affiliation{Department of Physics, University of Washington, Seattle, WA 98195-1560, USA}}

\AFFicrr
\AFFkashiwa
\AFFmad
\AFFbu
\AFFubc
\AFFbnl
\AFFuci
\AFFcsu
\AFFcnm
\AFFduke
\AFFfukuoka
\AFFgifu
\AFFgist
\AFFuh
\AFFicl
\AFFkek
\AFFkobe
\AFFkyoto
\AFFliv
\AFFmiyagi
\AFFnagoya
\AFFkmi
\AFFpol
\AFFsuny
\AFFokayama
\AFFosaka
\AFFox
\AFFqmul
\AFFregina
\AFFseoul
\AFFsheff
\AFFshizuokasc
\AFFstfc
\AFFskk
\AFFtokai
\AFFtokyo
\AFFtodai
\AFFipmu
\AFFtit
\AFFtus
\AFFtoronto
\AFFtriumf
\AFFtsinghua
\AFFynu
\AFFuw

\author{K.~Abe}
\AFFicrr
\AFFipmu
\author{C.~Bronner}
\author{G.~Pronost}
\AFFicrr
\author{Y.~Hayato}
\AFFicrr
\AFFipmu
\author{M.~Ikeda}
\AFFicrr
\author{K.~Iyogi}
\AFFicrr 
\author{J.~Kameda}
\AFFicrr
\AFFipmu 
\author{Y.~Kato}
\AFFicrr
\author{Y.~Kishimoto}
\AFFicrr
\AFFipmu 
\author{Ll.~Marti}
\AFFicrr
% ...

\collaboration{The Super-Kamiokande Collaboration}
\noaffiliation

\begin{abstract}

We present the results of a search in the Super-Kamiokande (SK) detector
for excesses of neutrinos with energies above a few GeV that are in
the direction of the track events reported in IceCube. Data from all
SK phases (SK-I through SK-IV) were used, spanning a period from April 1996 to
April 2016 and corresponding to an exposure of 225 kilotonne-years
. We considered the 14 IceCube track events from a data set with 1347
livetime days taken from 2010 to 2014. We use Poisson counting to
determine if there is an excess of neutrinos detected in SK in a 10
degree search cone (5 degrees for the highest energy data set) around
the reconstructed direction of the IceCube event. No significant
excess was found in any of the search directions we examined. We also
looked for coincidences with a recently reported IceCube multiplet
event. No events were detected within a $\pm$ 500 s time window around
the first detected event, and no significant excess was seen from that
direction over the lifetime of SK. 

\end{abstract}


\section{Introduction} \label{sec:intro}

Neutrino astronomy is a burgeoning field, bridging the gap between
astronomy and particle physics. Neutrinos travel undistorted from
their source, and are therefore a valuable probe of the inner workings of
astrophysical phenomena. The first definitive measurement of high energy
extragalactic neutrinos was made by the IceCube
experiment in 2013 \citep{aartsen13}, where they were able to reject
the atmospheric-neutrino-only hypothesis at greater than 4 $\sigma$.

Detecting neutrinos that are astrophysical in origin has raised many
questions: Where are these neutrinos coming from? What process is
creating them? There has not yet been any significant evidence to
suggest that these neutrinos are pointing to a particular region of
the sky. The many searches for counterparts to the neutrino signal
have been largely unsuccessful, including searches for coincidences
with photons from Fermi LAT \citep{peng17}, fast radio bursts
\citep{fahey16}, as well as a search for coincidences with a large
catalogue of candidate sources \citep{aartsen17}. One search
\citep{kadler16} found a coincidence between a PeV
neutrino detected in IceCube and an outburst of the blazar PKS
B1424−418, giving a hint at a possible origin of these astrophysical
neutrinos.

IceCube uses detected events with energies above a few hundred TeV to
look for astrophysical neutrinos. In this energy region, the
atmospheric neutrino background is expected to be low. However, now
that astrophysical neutrino candidates have been identified, one can
search in the lower energy data for an excess of events
coming from the same direction. Given the unknown origin of these
neutrinos, searching this previously unexplored energy region is
of interest.

Super-Kamiokande (SK), a water Cherenkov detector located in Japan,
detects atmospheric neutrinos in the energy range of 30 MeV to several
TeV. For events with energies above a few GeV, the direction of
the incoming neutrino can be reconstructed as it is well correlated
with the direction of the detected outgoing lepton.

SK has performed a number of astrophysical neutrino searches in the past. Recent
searches include a general, all-sky astrophysical search \citep
{thrane09b, abe06, swanson06}, searches for
coincidences with gamma ray bursts, supernova remnants, and other
potential sources of astrophysical neutrinos \citep{thrane09a,
  desai08, thrane09b}, dark matter searches
\citep{tanaka11, choi15}, and a search for coincidences with
the recent detection of gravitational wave signals \citep{abe16}.

In this paper, we look for excesses of neutrino events in
Super-Kamiokande in the direction of the IceCube events from their
data release in \citet{aartsen15}. We use the IceCube events that have
the best pointing accuracy, known as track events, and determine if
there is an excess of events in the full SK high energy dataset. Given
the uncertainty about the origins of these astrophysical neutrinos, we
perform a model-blind search, without assuming an energy spectrum a
priori. Since we have no observational or theoretical motivation for the
time duration over which these neutrinos are emitted, we do not
require any timing correlation between the IceCube and SK events. For
this simple estimate, we omit discussions of systematic errors. We
also report the search for coincidences with the recent multiplet
event reported in \citet{aartsen17}.

An estimate of the number of events anticipated in the SK sample can
be derived by extending the point source limits from \citet{aartsen16}
down to lower energies. Assuming the standard E$^{-2}$ spectrum, we
would expect approximately 0.5 events in the UPMU sample and
approximately 0.002 events in FC
and PC samples. If we assume a cutoff of E$_{\nu} <$ 100 TeV (which is below the
deposited energies of IceCube events 4, 9, and 12), we would expect
around 5
events in the UPMU sample and around 0.02 events in the FC and PC
samples. More events could be expected in the SK topologies if there
is a softer spectral index (in \citet{aartsen15} the best fit index is
−2.58 $\pm$ 0.25) or if there is a broken power law between the higher
energy domain described by IceCube and the lower energies seen in SK.

\section{The Super-Kamiokande Experiment}

The SK detector is a 50-kilotonne (22.5 kilotonne fiducial) water
Cherenkov detector located in the Mozumi mine in the Gifu prefecture
of Japan. The cylindrical detector is optically separated into an
inner detector (ID) volume, which is viewed by $\sim$ 11,000
photomultiplier tubes (PMTs), and an outer detector (OD) volume, which
is viewed by $\sim$ 2,000 PMTs. A more detailed description of the
SK detector can be found in \citet{fukuda03}.

The SK data is divided into four experimental phases. SKI ran from
1996-2001 with 40\% photocoverage. In 2001, there was an accidental
implosion that damaged some of the PMTs. SKII ran from 2002-2005 with a
photocoverage of 20 \%. In 2006, the damaged phototubes were repaired
and the SKIII phase began with 40\% photocoverage. After an
electronics upgrade in 2008, the current phase of the experiment,
SKIV, began. Data from all four phases of the experiment are used in
this analysis, spanning April 1996 - April 2016 and corresponding to
225 kilotonne-years.

Detected neutrino events at energies above 30 MeV can have three
different topologies in the SK detector. The first, known as
fully-contained (FC) events, have a reconstructed vertex inside the
fiducial volume, with little light seen in the OD. Events that have a
reconstructed vertex inside the fiducial volume, but have interaction
products that produce light in the OD, are known as
partially-contained (PC) events. Finally, muon neutrinos that interact
in the surrounding rock and produce penetrating muons are known as
upward-going muon (UPMU) events. We require these events to be coming
from below the horizon in order to distinguish them from cosmic
muons. These topologies roughly represent increasing, though
overlapping, energy regions. For the atmospheric neutrino energy
spectrum, FC events have an average energy of about
1 GeV, PC events have an average energy of about 10 GeV, and UPMU events have
an average energy of about 100 GeV. More information on the SK topologies,
including the selection cuts used in the data reduction, can be found
in \citep{ashie05}.

\section{Search Method} \label{sec:searchmethod}

The IceCube search directions were taken from the October 2015 data release
\citep{aartsen15}. This data set contains neutrino candidates with two
different topologies: track and shower. Track events are mainly from
muon neutrino charged-current interactions, while shower events are
mainly from neutral current interactions, as well as electron and tau neutrino
  charged-current interactions. Only IceCube track events were used in this
analysis. Shower events typically have poor angular resolution, with
an error up to 20$^{\circ}$ for the \citep{aartsen15} data set, making
them unsuitable for a coincidence analysis using only spatial information. Track
events, on the other hand, have a median angular resolution of better than
1$^{\circ}$, allowing us to perform a spatial coincident
search. Table \ref{Table:IC_event_info} lists the properties of the
IceCube events.

\begin{table}[hptb]
\centering
\begin{tabular}{c|c|c|c}  \label{Table:IC_event_info}
Event Number & Declination (degrees) & RA (degrees) & Deposited Energy (TeV) \\
\hline
1 &  -31.2 & 127.9 & 78.7 \\
2 &  -0.4 & 110.6 & 71.4 \\
3 &  -21.2 & 182.4 & 32.6 \\
4 &  40.3 & 67.9 & 252.7 \\
5 &  -24.8 & 345.6 & 31.5 \\
6 &  -13.2 & 208.7 & 82.2 \\
7 &  -71.5 & 164.8 & 46.1 \\
8 & 20.7 & 167.3 & 30.8 \\
9 & 14.0 & 93.3 & 200.5 \\
10 & -22.0 & 206.6 & 46.5 \\
11 & 0.0 & 336.7 & 84.6 \\
12 & -86.3 & 219.0 & 429.9 \\
13 & 67.4 & 209.4 & 74.3 \\
14 & -37.7 & 239.0 & 27.6 \\
\end{tabular}
\caption{Information on the track events from IceCube used in this
  coincidence search. The data were taken from \citet{aartsen15}.}
\end{table}

To ensure that the detected lepton points back to the incoming
neutrino, a low energy threshold was imposed on the FC and PC
datasets. A minimum energy requirement was determined by
calculating the lowest energy such that 50\% of the reconstructed
lepton directions of that energy agreed to within 10$^\circ$ of the
incoming neutrino direction in the simulated data set from our Monte
Carlo (MC) code. This threshold was determined to be 3.8 GeV for FC
events and 2.1 GeV for PC events. No explicit upper energy cut was
applied. 

A search cone, centered at the reconstructed direction of each IceCube
event, was defined with a half-angle opening of 10$^\circ$ for FC and
PC events and 5$^\circ$ for UPMU events. The UPMU events are higher in
energy than the other topologies and therefore the detected lepton
points back to the incoming neutrino with more accuracy, allowing for
a smaller search cone. 

In this analysis, only basic selection cuts were applied
post-reduction. For the FC data set, the cuts ensured that the
reconstructed event vertex was more than 2.0 m from the detector wall,
that the visible energy in the detector was greater than 3.8 GeV, and
that there were fewer than 16 hits (10 for SK-I) in the outer detector
volume. For the PC data set, the cuts ensured that the reconstructed
event vertex was more than 2.0 m from the detector wall, that the
visible energy in the detector was greater than 2.1 GeV, and that
there was greater than or equal to 16 hits (10 for SK-I) in the outer
detector volume. Finally, the UPMU data selection cuts ensured that
the fit direction was below the horizon, that the fit track length was
greater than 7.0 m for the events that the fitter classified as passing
through the detector, and that the fitted momentum of the lepton was
greater than 1.6 GeV for events that the fitter classified as stopping
in the detector.

To determine if there is an excess of events coming from the direction
of the high energy IceCube track events, Poisson statistics were
used. A test statistic was constructed based on the maximum likelihood
method. The likelihood of seeing N events in our search cone given the
expected number of background events including oscillations ($N_B$)
is,

\begin{equation}\label{eqn:likelihood}
L = \frac{e^\frac{-N_B}{\left(1-\alpha\right)}}{N!}\left(\frac{N_B}{1-\alpha}\right)^N,
\end{equation} 

\noindent where $\alpha$ represents the fraction of the N events that are from
signal and is the parameter over which we maximize.

The background to this search is from atmospheric neutrinos. The
SK MC code, along with a scaling for neutrino oscillations and the
overall normalization of the flux, was used to determine the number of
background events ($N_B$) we expect to see in our search cone. The MC code used Geant3
\citep{brun87} to simulate particle interactions and tracking. We used 500 years of
simulated atmospheric neutrino events for each SK phase. The truth
information was generated using nuclear interaction models used in
NEUT \citep{hayato09}. We scaled the MC sample for each phase to the
appropriate livetime, and then summed them together. Events were
assigned right ascensions assuming a flat local sidereal time, and so
the resulting $N_B$ was assumed to depend only on declination. The
combined sample was then scaled on an event-by-event basis to the
all-sky, best-fit value from data, which accounts for the flux
normalization and oscillations.

Our test statistic, $\Lambda$, is then, 

\begin{equation} \label{eqn:lambda}
\Lambda = 2\log\frac{L(\alpha_{\mathrm{fitted}})}{L(\alpha=0)},
\end{equation}

\noindent where $\alpha_{\mathrm{fitted}}$ is obtained from maximizing Equation
\ref{eqn:likelihood}. This is the final indicator of the statistical
significance for the number of measured neutrino events in our search
cone.

\section{Results} \label{sec:results}

\begin{figure}[hptb] 
\centering
\includegraphics[trim=3cm 1cm 3.5cm 1.5cm, clip=true, width=0.28\linewidth]{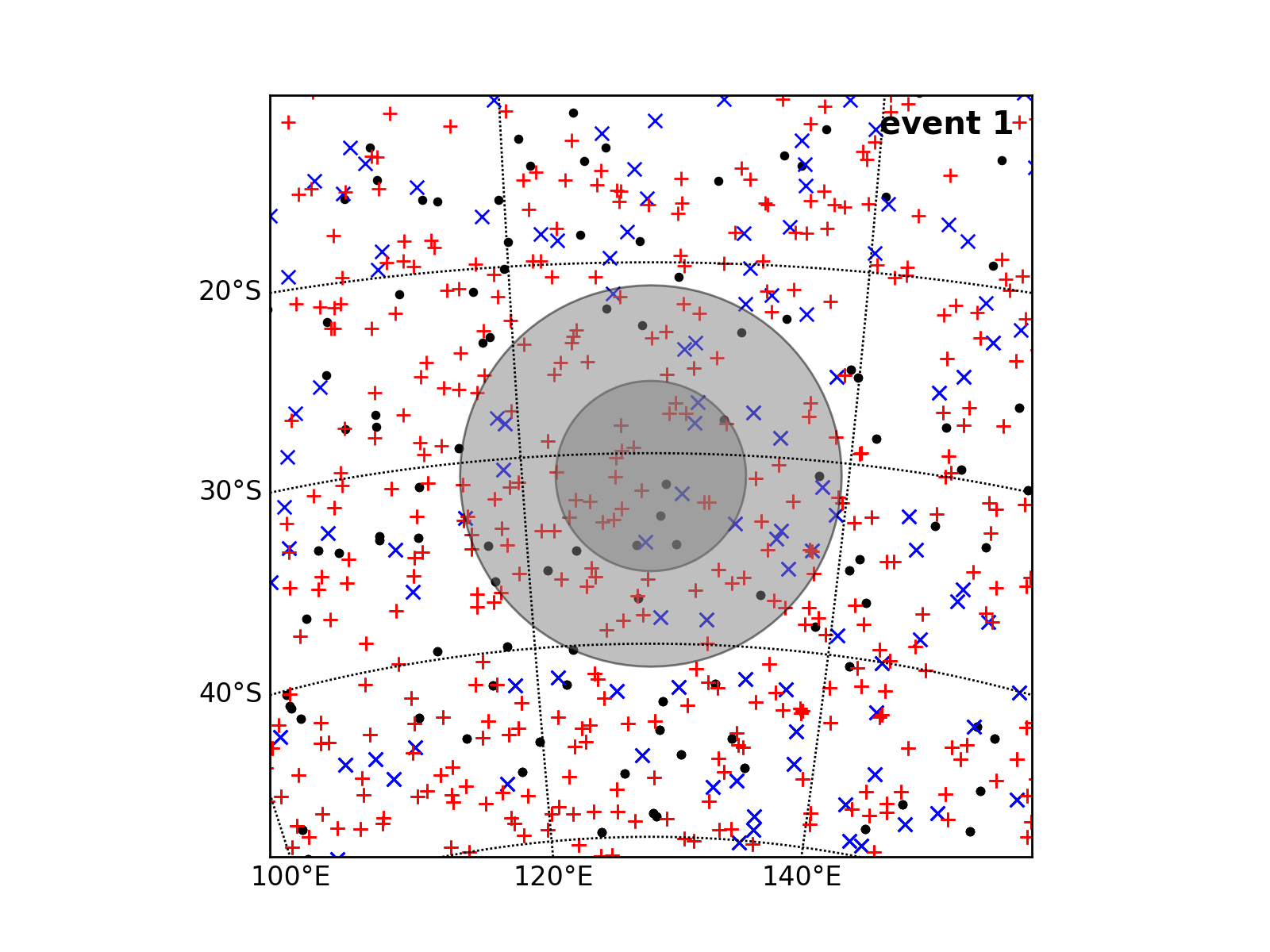} 
\includegraphics[trim=3cm 1cm 3.5cm 1.5cm, clip=true, width=0.28\linewidth]{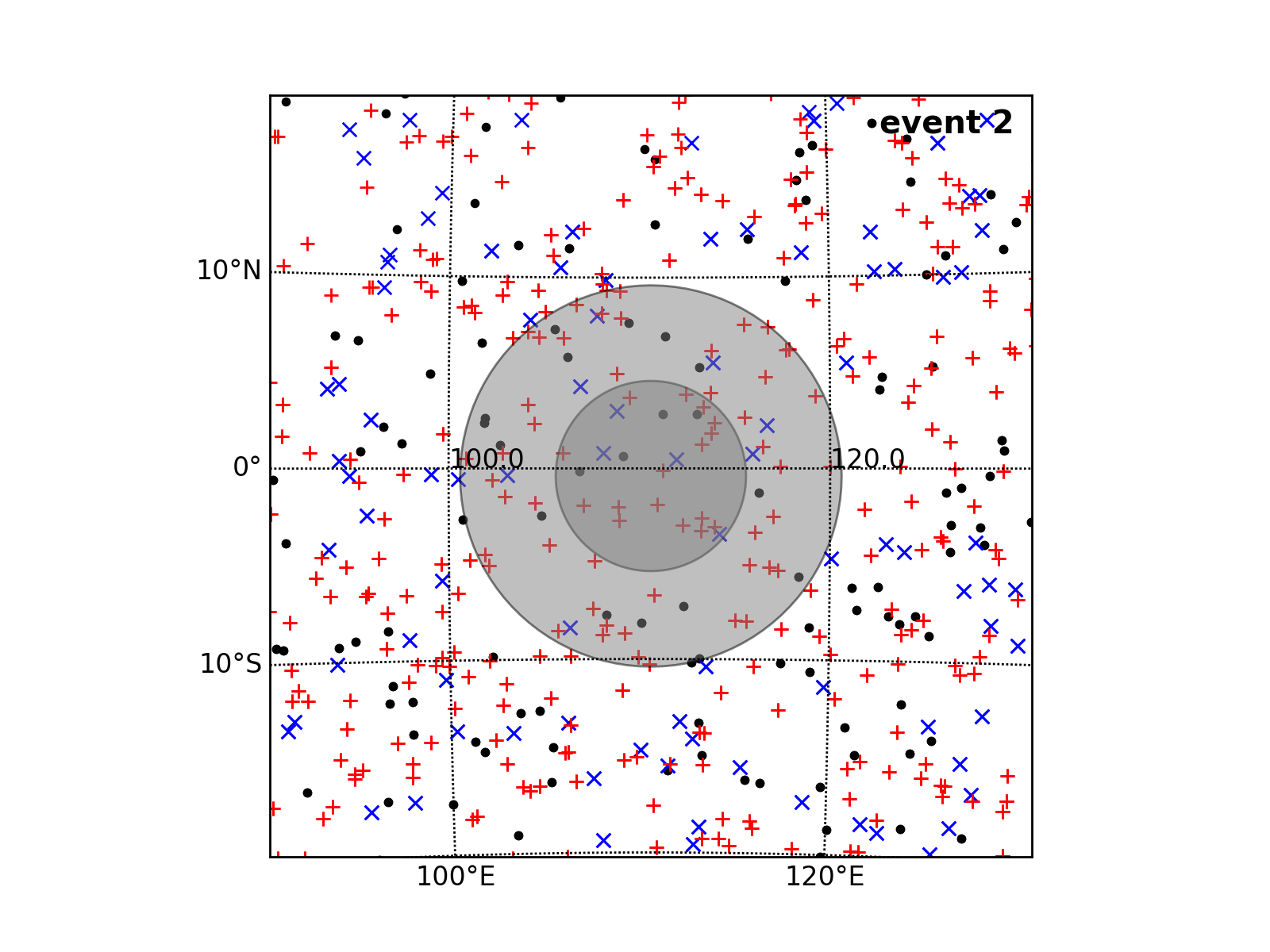} 
\includegraphics[trim=3cm 1cm 3.5cm 1.5cm,clip=true,width=0.28\linewidth]{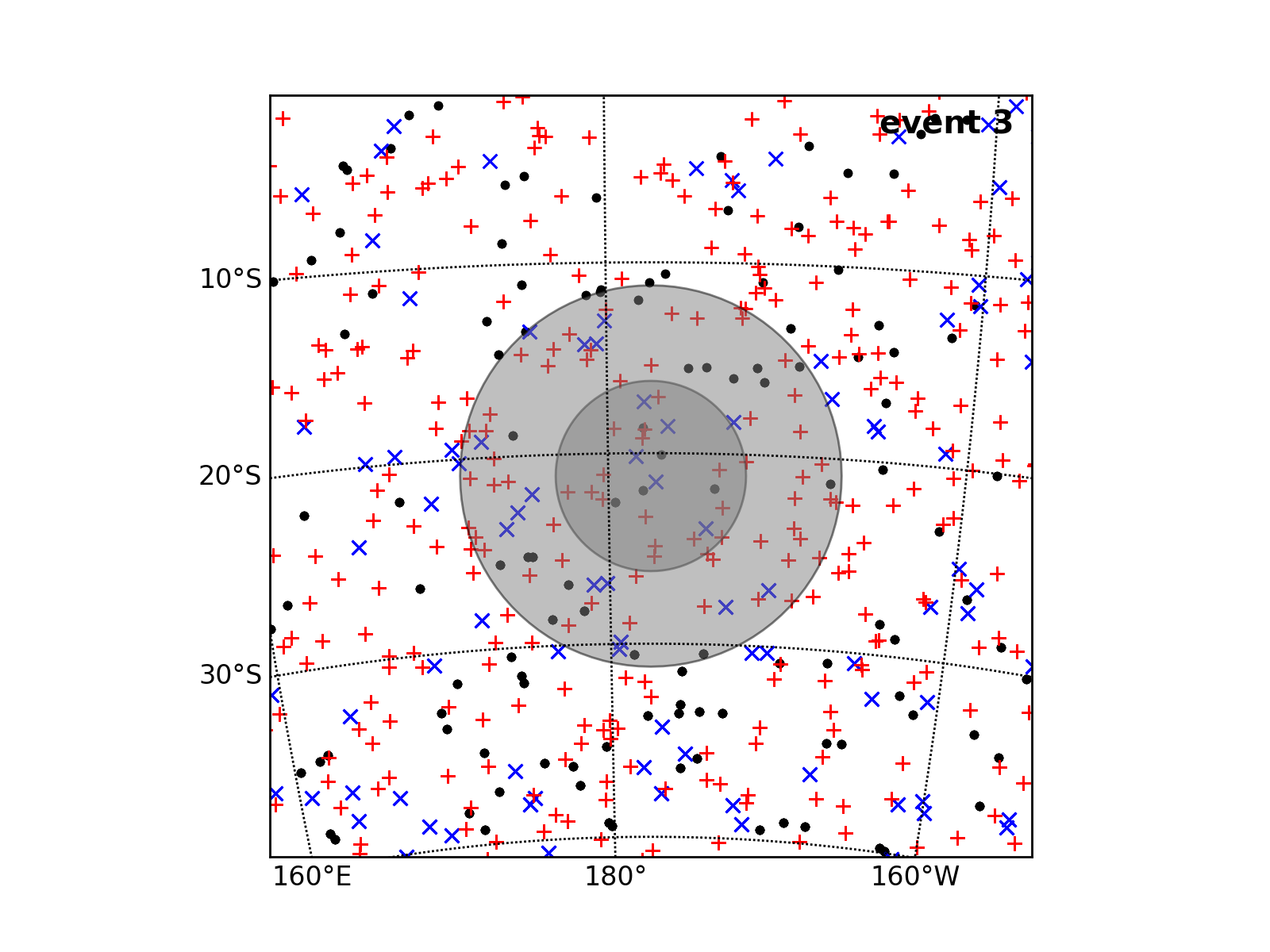}\\
\includegraphics[trim=3cm 1cm 3.5cm 1.5cm, clip=true, width=0.28\linewidth]{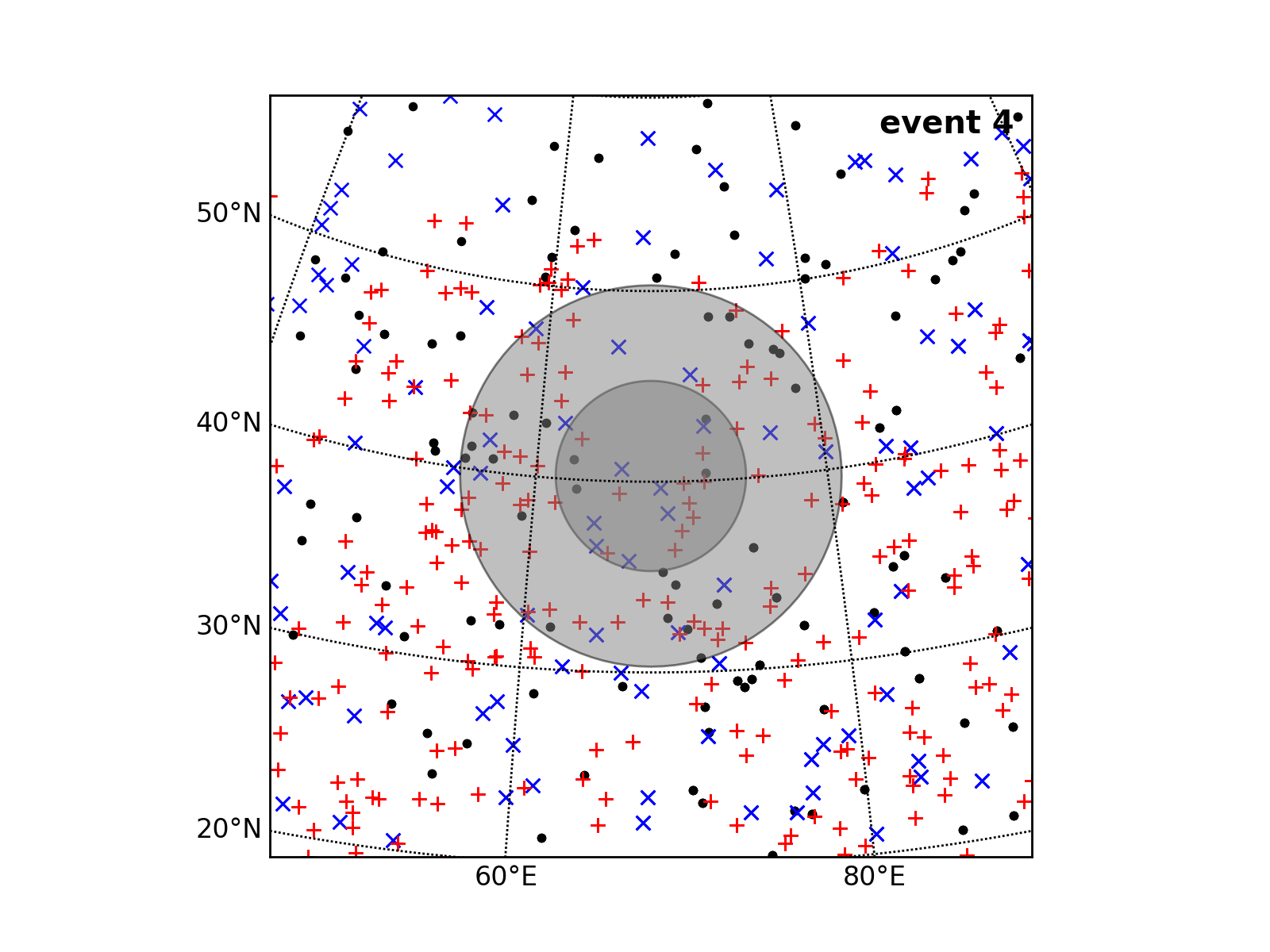} 
\includegraphics[trim=3cm 1cm 3.5cm 1.5cm, clip=true, width=0.28\linewidth]{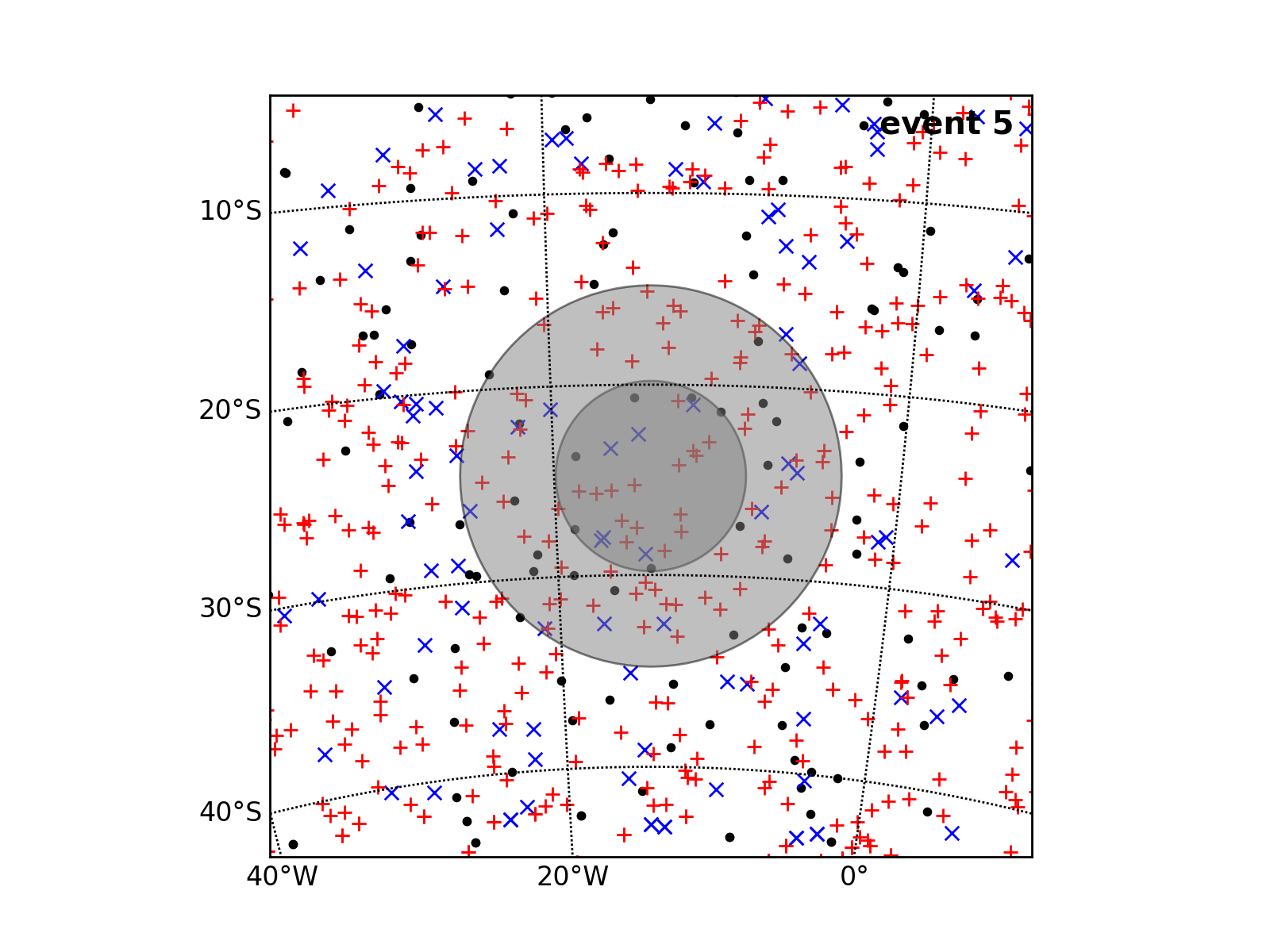} 
\includegraphics[trim=3cm 1cm 3.5cm 1.5cm, clip=true, width=0.28\linewidth]{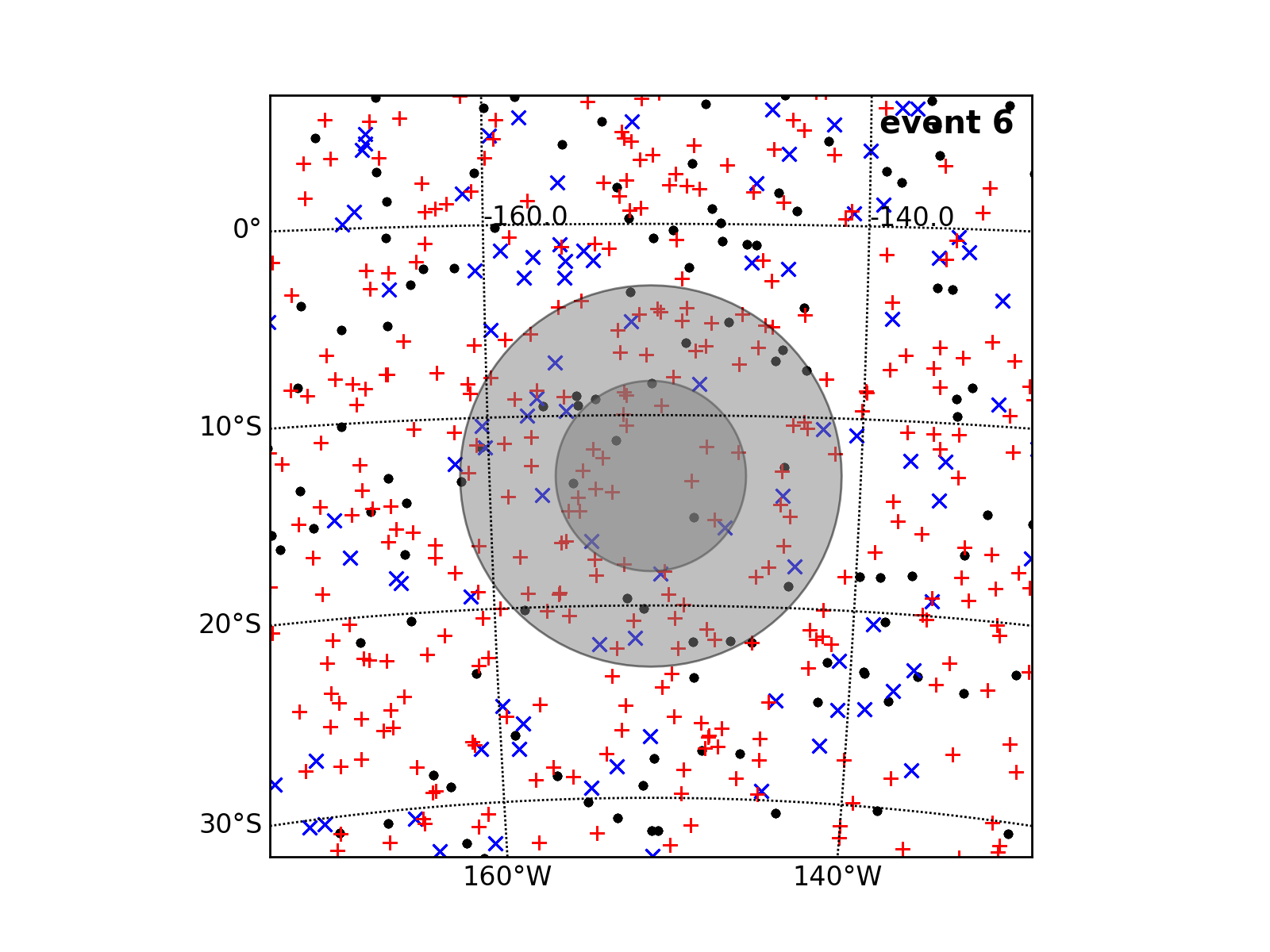} \\
\includegraphics[trim=3cm 1cm 3.5cm 1.5cm, clip=true, width=0.28\linewidth]{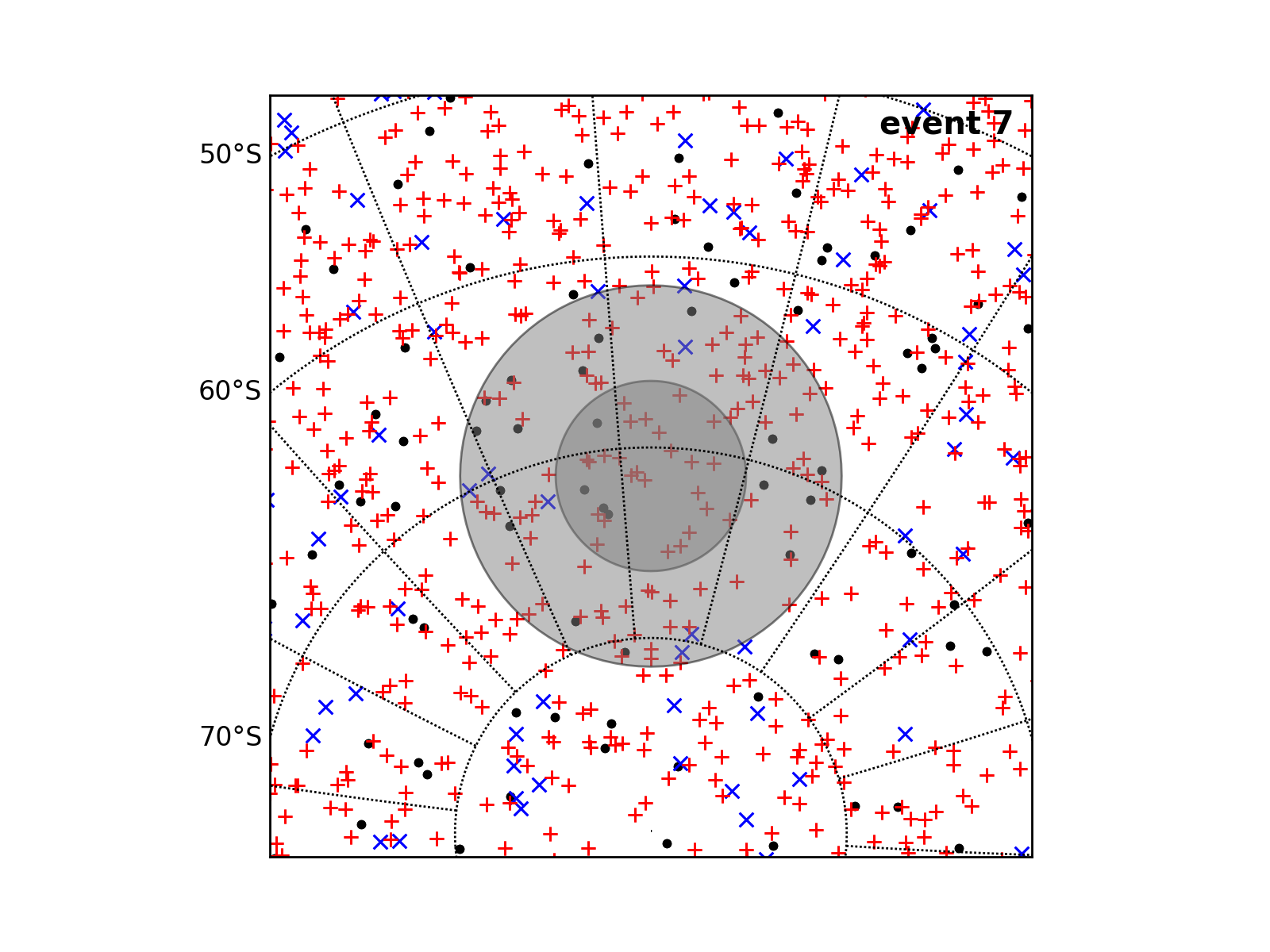} 
\includegraphics[trim=3cm 1cm 3.5cm 1.5cm, clip=true, width=0.28\linewidth]{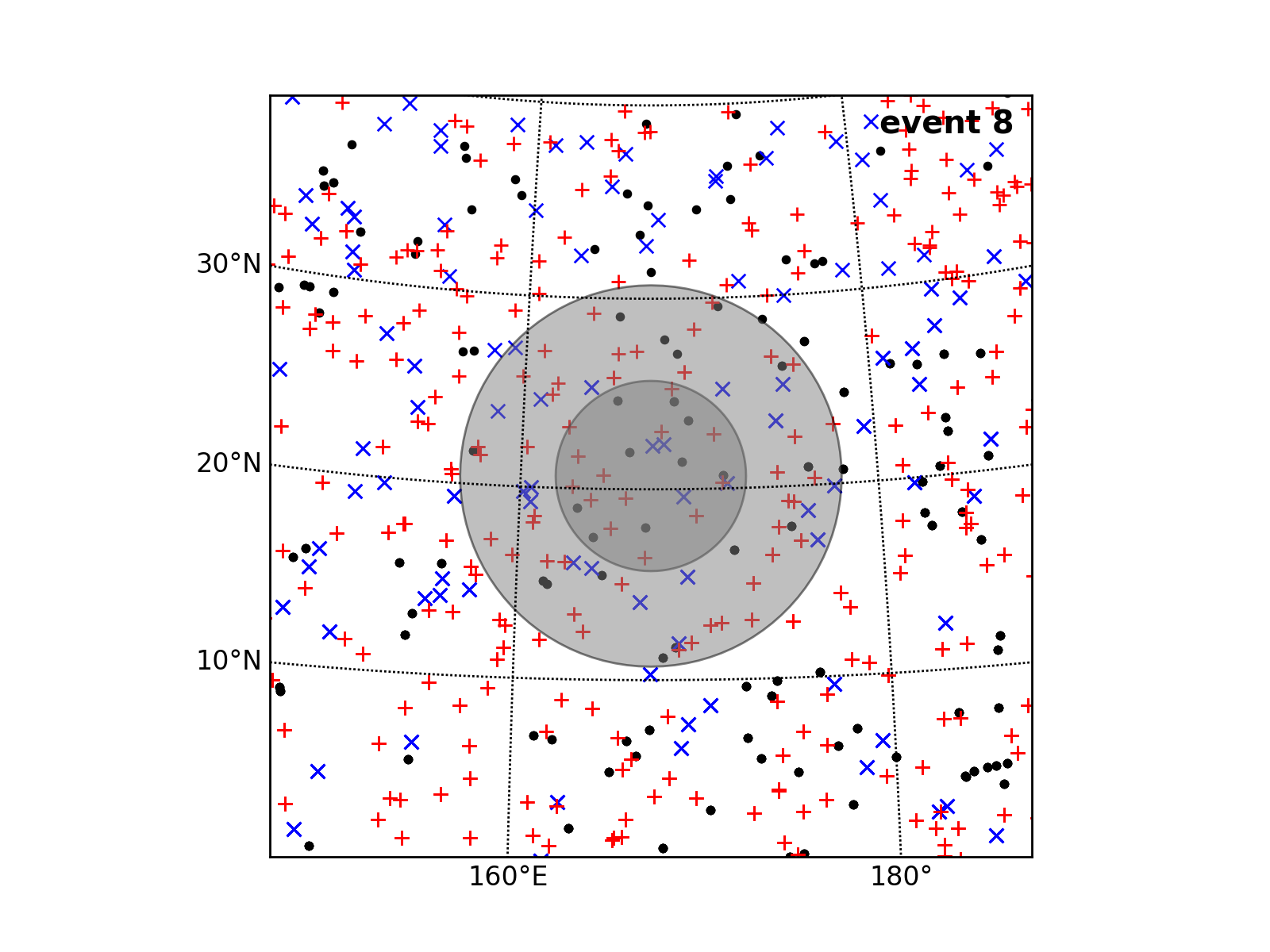} 
\includegraphics[trim=3cm 1cm 3.5cm 1.5cm, clip=true, width=0.28\linewidth]{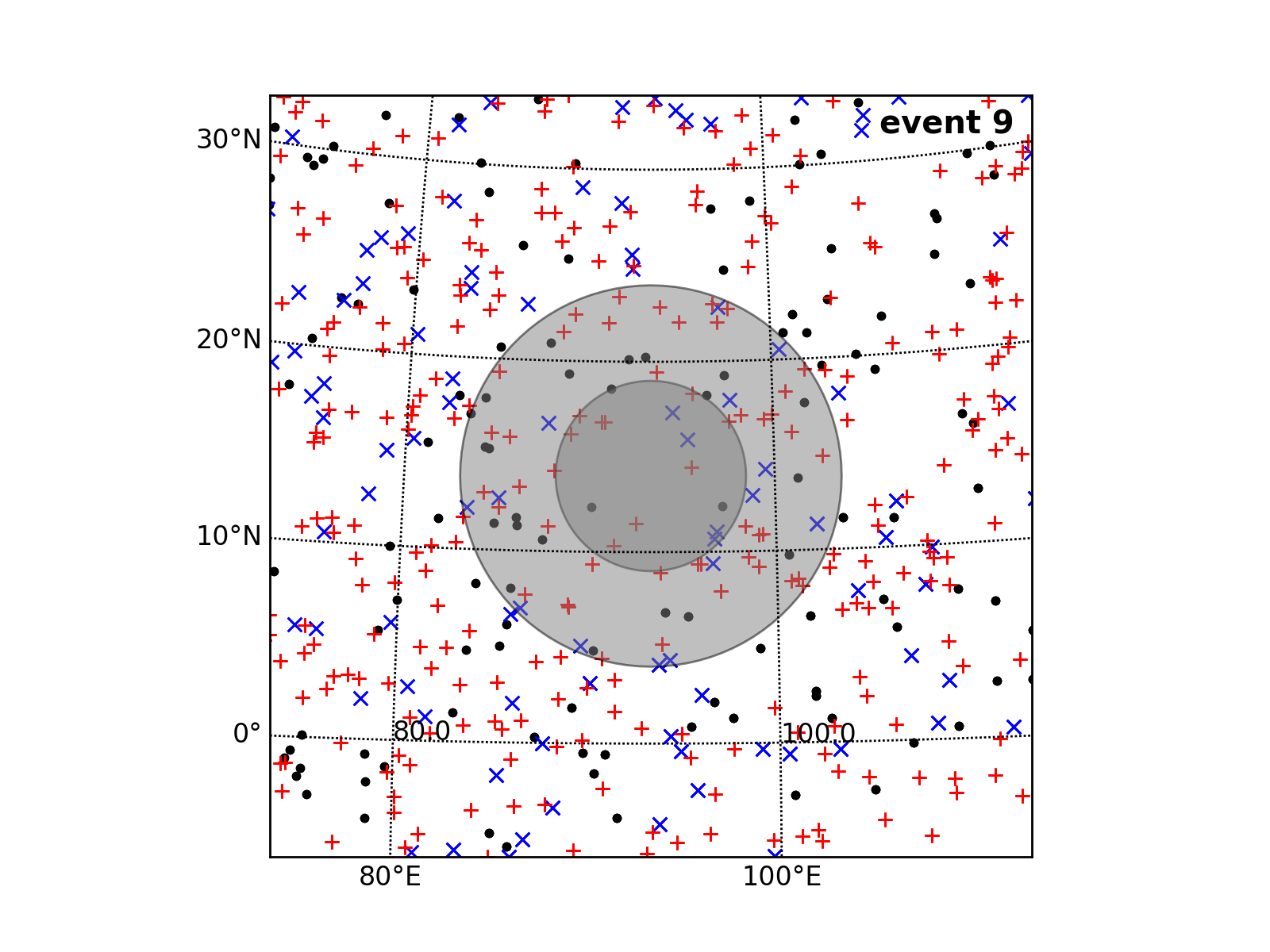} \\
\includegraphics[trim=3cm 1cm 3.5cm 1.5cm, clip=true, width=0.28\linewidth]{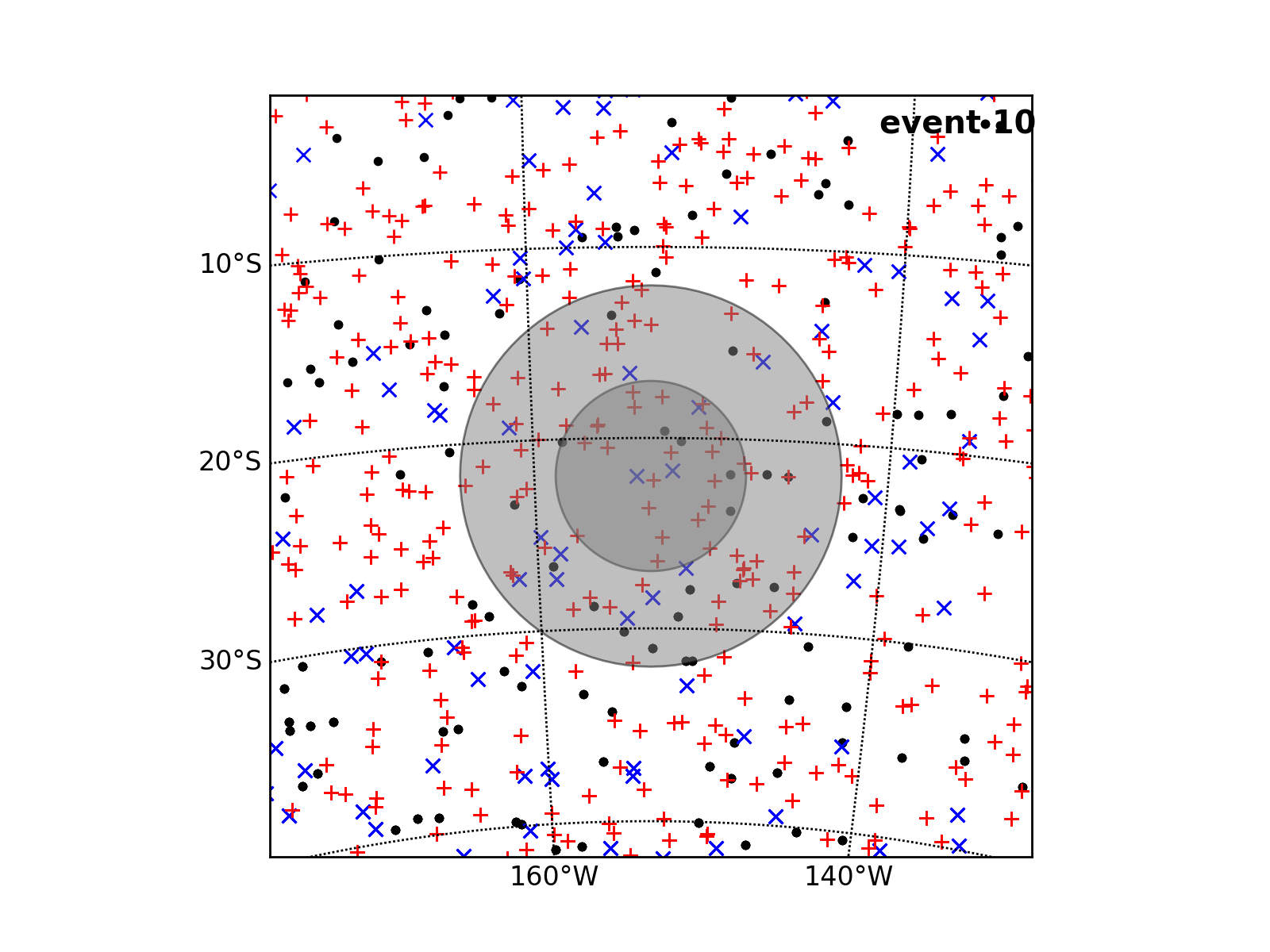} 
\includegraphics[trim=3cm 1cm 3.5cm 1.5cm, clip=true, width=0.28\linewidth]{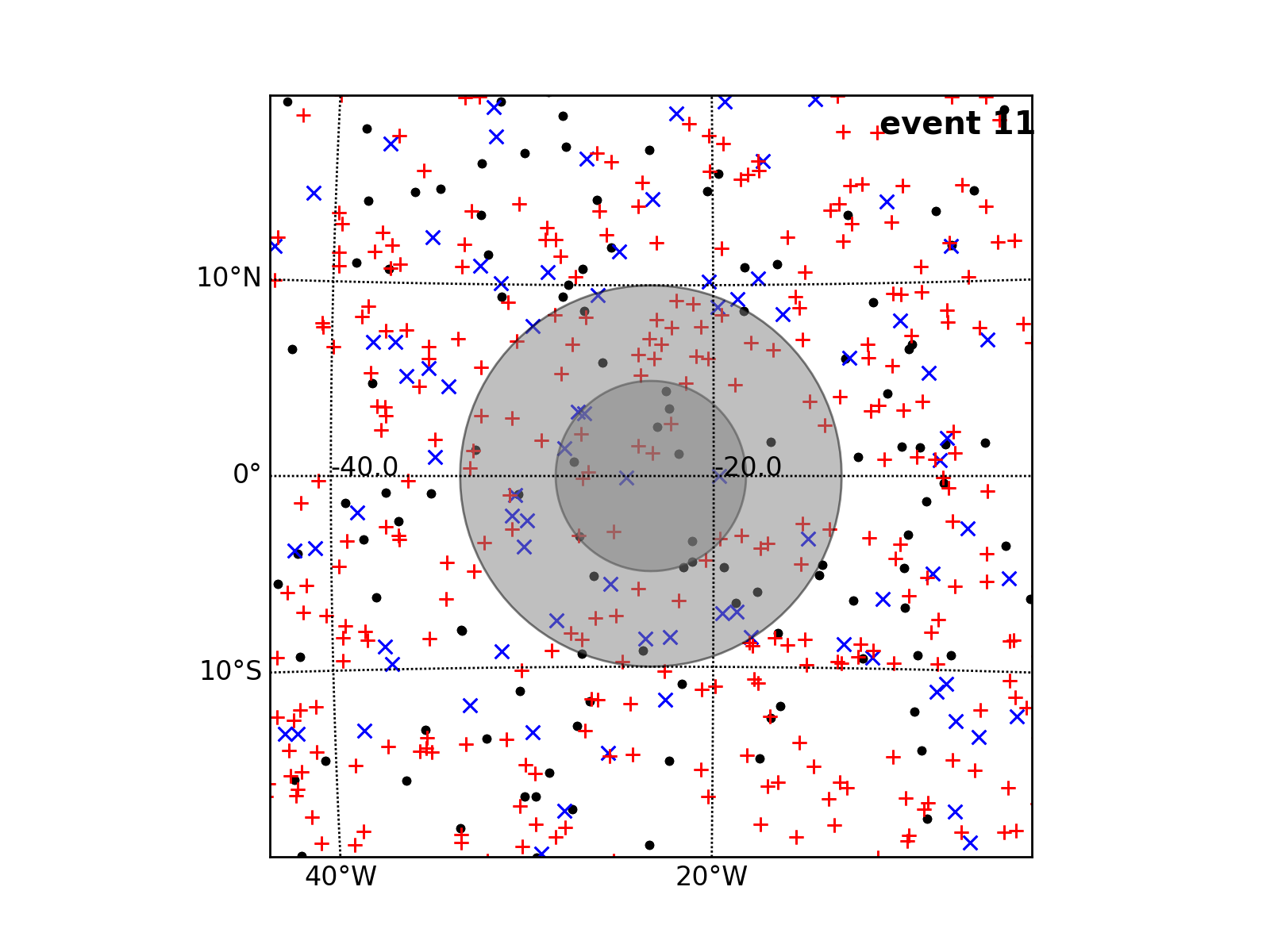} 
\includegraphics[trim=3cm 1cm 3.5cm 1.5cm, clip=true, width=0.28\linewidth]{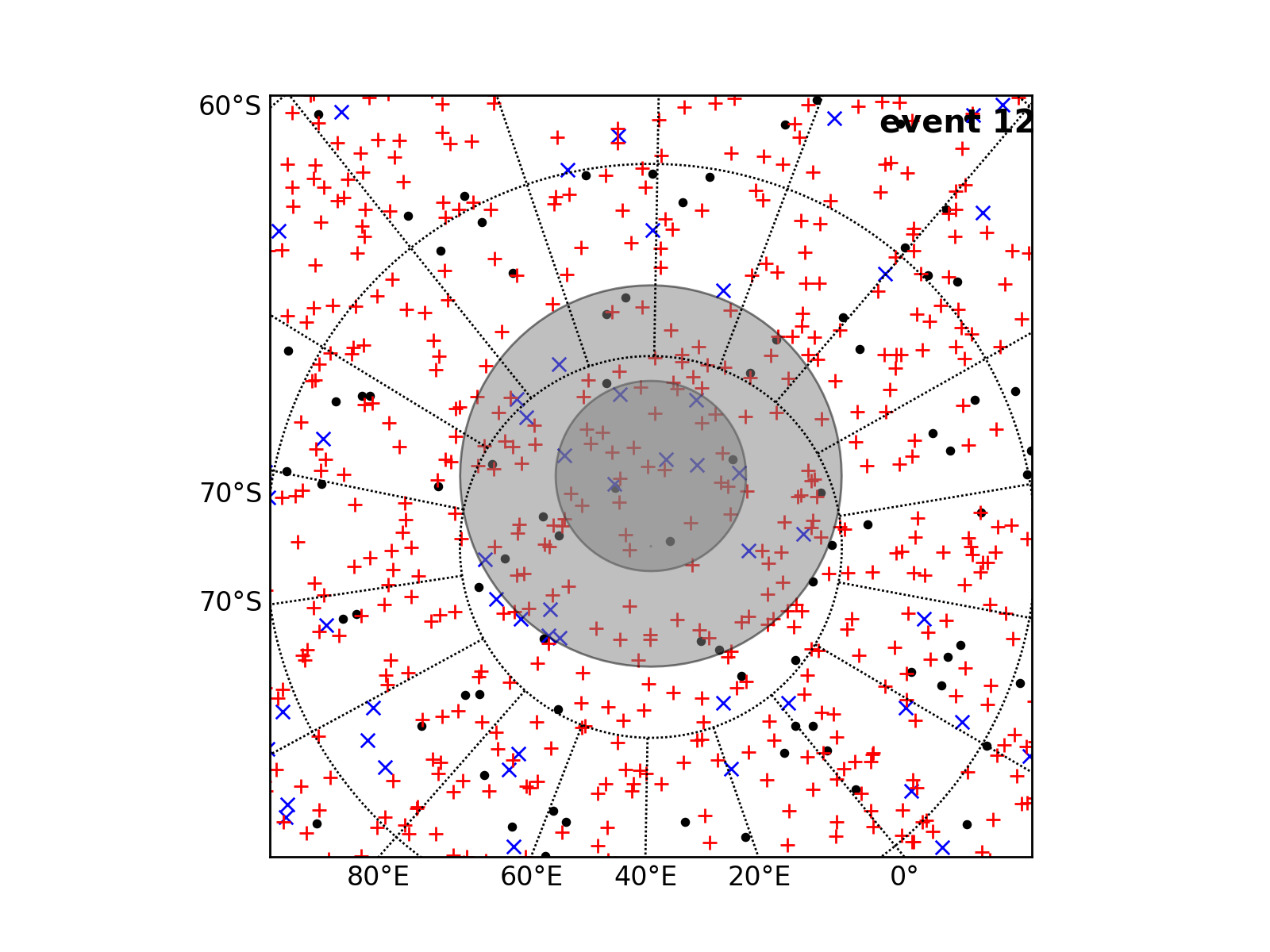}\\ 
\includegraphics[trim=3cm 1cm 3.5cm 1.5cm, clip=true, width=0.28\linewidth]{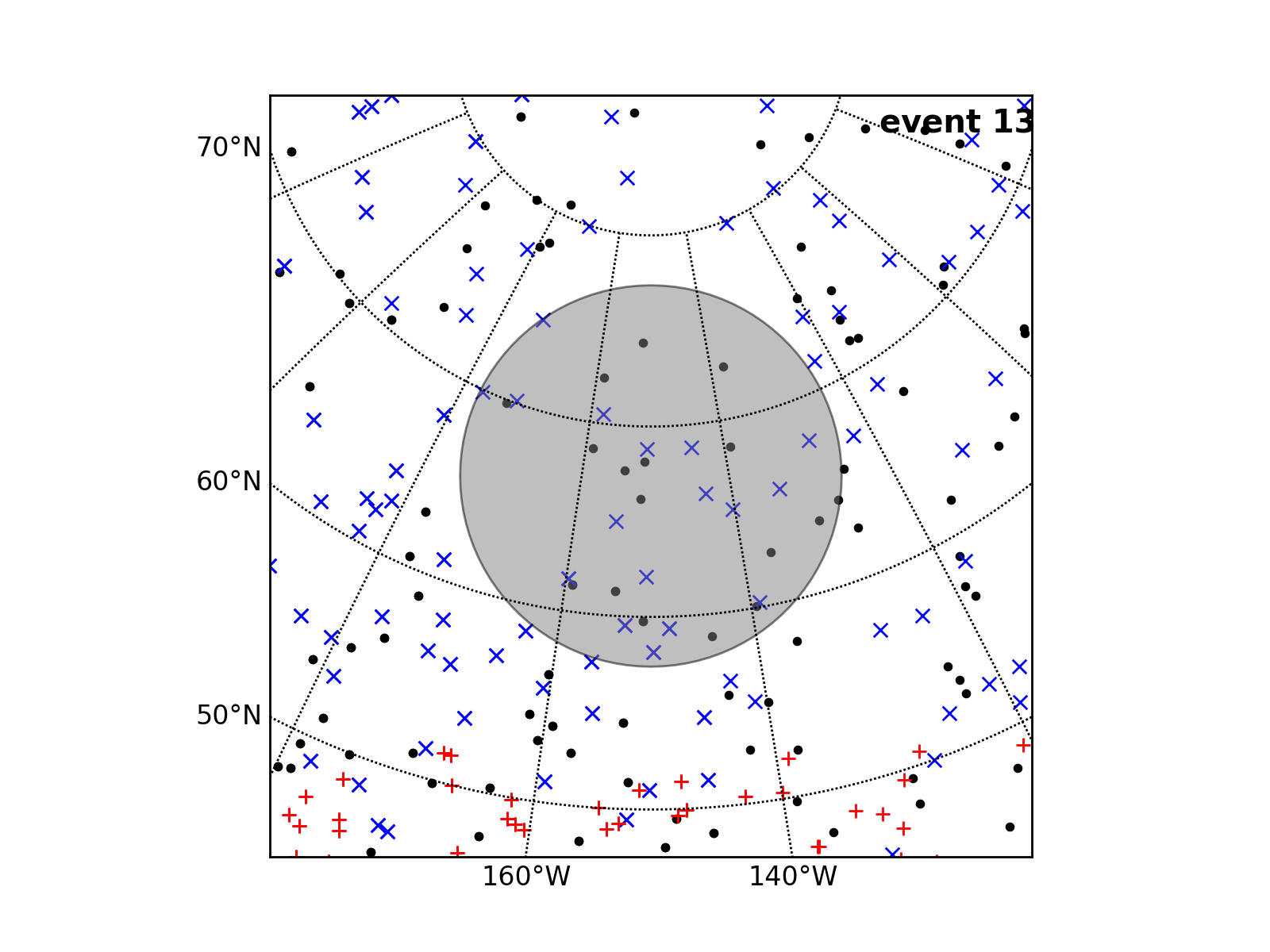} 
\includegraphics[trim=3cm 1cm 3.5cm 1.5cm, clip=true, width=0.28\linewidth]{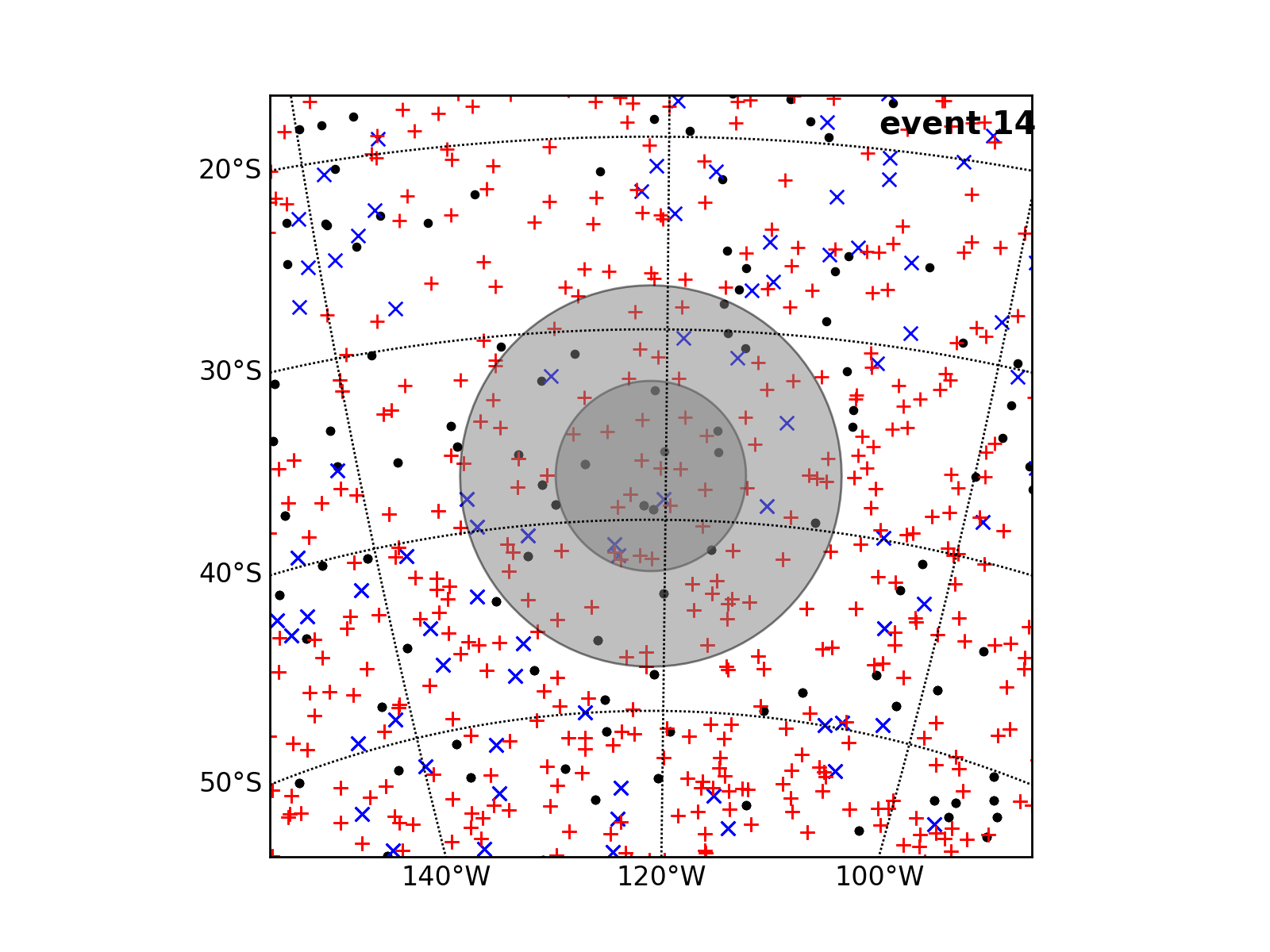} 
\caption{The detected FC (black circle), PC (blue x) and UPMU (red
  +) events in and around the search window. The position of events are shown in
  equatorial coordinates with right ascension on the x-axis and
  declination on the y-axis. The dark grey disk
  shows the 5 degree search cone used for UPMU events, while the light
grey disk shows the 10 degree search cone used for FC and PC
events.}\label{fig:scatterplot}
\end{figure}

Figure \ref{fig:scatterplot} shows the spatial distribution of the
detected neutrino events in the region of each search direction. The
density of the events are dependent on declination. This can be seen
most clearly in the UPMU data sample, where there is a high density
near the southern pole and no events at declinations above 54$^\circ$
(see event 13 in Figure \ref{fig:scatterplot}).
The density of detected neutrino events does not appear to be
significantly higher inside any of the search regions compared with
the density around the search regions.

\begin{figure}[hptb] 
\includegraphics[trim=0cm 0cm 1.5cm 0cm, clip=true, width=0.33\linewidth]{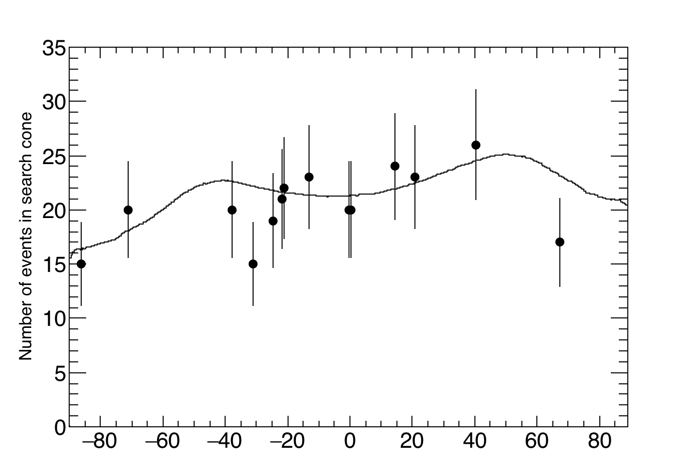} 
\includegraphics[trim=0cm 0cm 1.5cm 0cm, clip=true, width=0.33\linewidth]{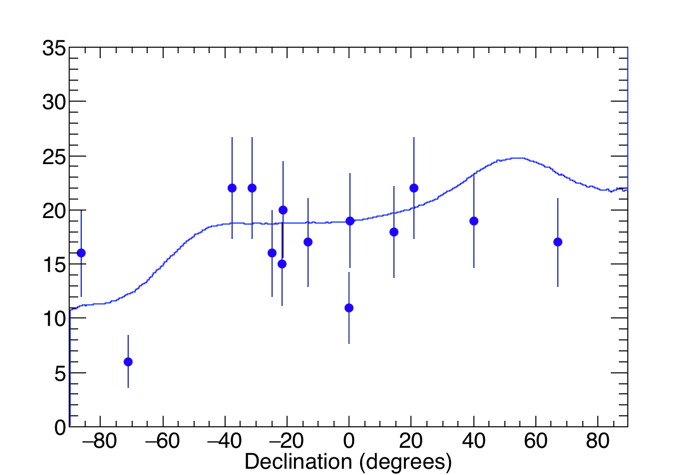} 
\includegraphics[trim=0cm 0cm 1.5cm 0cm, clip=true, width=0.33\linewidth]{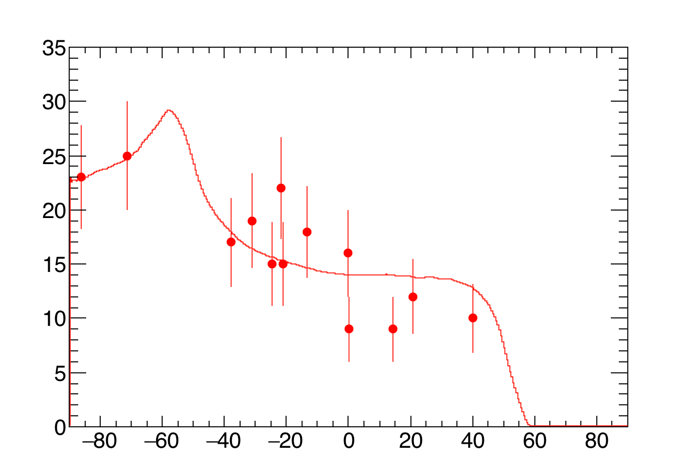} 
\caption{The number of events expected in the search cone
  (solid line) for FC (black, left), PC (blue, center), and UPMU (red,
  right) topologies,
  shown with the number of events found in the search cone from the
  data (points). The errors shown here are $\sqrt{N}$. Event 13
  (declination of 67.4$^{\circ}$) was not visible to the UPMU data set
since it is always above the horizon in the SK detector.}\label{fig:backgrounds}
\end{figure}

Figure \ref{fig:backgrounds} shows the background expectations for the
three topologies considered in this search. The background expectation
is assumed to be independent of the right ascension. The FC and PC topologies
extend to all declinations and have a slightly positive slope due to
oscillations in the upward-going neutrinos. The two peaks are due to
the atmospheric neutrinos coming from near the horizon. Here, the
cosmic rays are more likely to interact due to the fact that
the atmosphere is thicker and that path length for traversing
this region is longer and so there is a greater chance for the cosmic
ray to decay. The UPMU topology requires that the neutrino events come
from below the horizon, and thus there are no events where the
reconstructed direction has declinations
above 54 $^{\circ}$. Neutrinos coming from declinations of greater than
-54$^{\circ}$ spend increasingly more time above the horizon, and so there is a decreasing trend above this
declination. The maximum number of UPMUs are at -54$^{\circ}$. This is
because these neutrinos are near, but always below the horizon, where
a greater flux of atmospheric neutrinos is expected.

We then used the SK data to calculate the number of detected events
($N$) in the search cone for each of the IceCube search directions.
Figure \ref{fig:backgrounds} shows the number of detected neutrino
events in the search cone compared with the background expectation. 

To determine if there were any statistically significant excesses in
our data, we calculated the test statistic, $\Lambda$, for each search
direction using Equation \ref{eqn:lambda}. The expected distribution
of $\Lambda$ was also calculated using the 500-year simulated MC data
set. The UPMU topology was used and the declination of -31.2$^{\circ}$
was assumed. The events were first randomly assigned a right ascension
assuming a flat local sidereal time. The number of events in a
5$^{\circ}$ search cone was then determined, scaled for the
appropriate livetime. This was compared to the background expectation
$N_{B}$ for this declination and $\Lambda$ was calculated using
Equation \ref{eqn:lambda}. This algorithm was repeated 1$\times 10^7$
times, randomly assigning new right ascension values to the data each
time. The expected distributions for the different topologies, as well
as the different declinations, were checked and found to be the same.

\begin{figure}[hptb] 
\centering
\includegraphics[trim=0cm 0cm 1.5cm 1.cm, clip=true, width=0.55\linewidth]{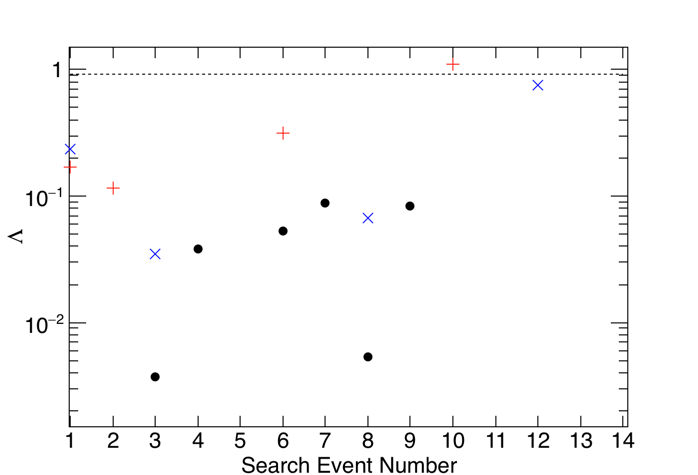} 
\caption{The test statistic for each search direction, plotted for the
  FC (black circle), PC (blue x), and UPMU (red +) topologies. The
  dashed line represents the 1 $\sigma$ expectation calculated from
  the atmospheric MC. No trend can be seen for an excess in a particular
search direction.}\label{fig:lambda_results_all}
\end{figure}

Figure \ref{fig:lambda_results_all} shows the test statistic for all
three topologies for each search direction. $\Lambda$ was found to be
zero most often, signifying that the number of events in the search
cone best fit to the expected background, or that there were fewer
events in the search cone than the expected background would
suggest. No excess of greater than 2$\sigma$ was found for any of the
topologies in any of the search directions. No search direction
jointly yielded a significantly higher $\Lambda$ in all three of the
topologies.

The most significant event had a calculated $\Lambda$ of 1.1, which
corresponds to a significance of about 1.1 $\sigma$ from the MC
background prediction. This was in the UPMU data sample in
the direction of event number 10, corresponding to a declination of
-22$^{\circ}$. In this search direction, we observed 22 events and
expected 13.7 events from the atmospheric background MC prediction. 

\begin{figure*}
\gridline{\fig{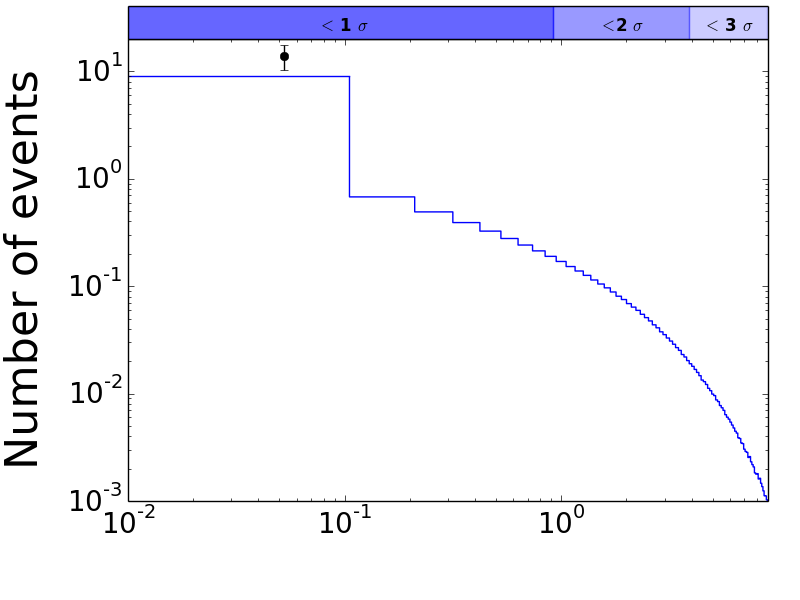}{0.33\textwidth}{(a)}
          \fig{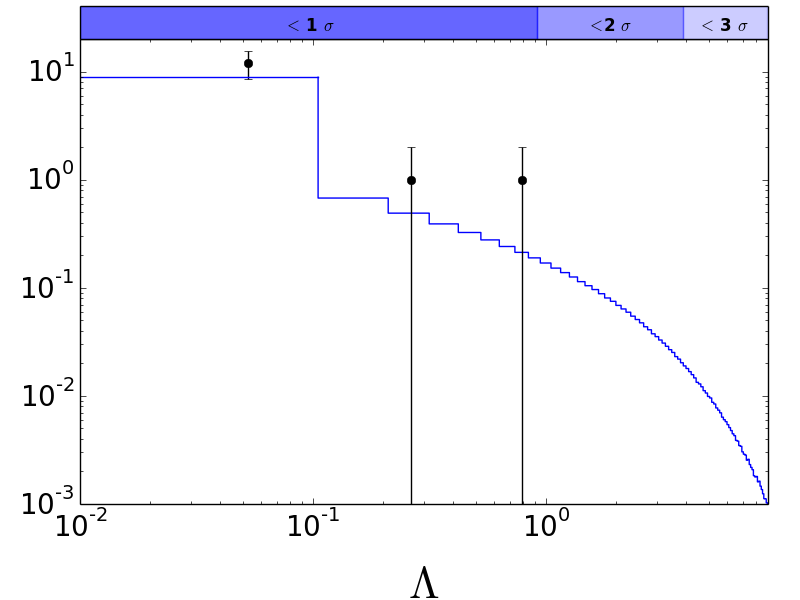}{0.33\textwidth}{(b)}
          \fig{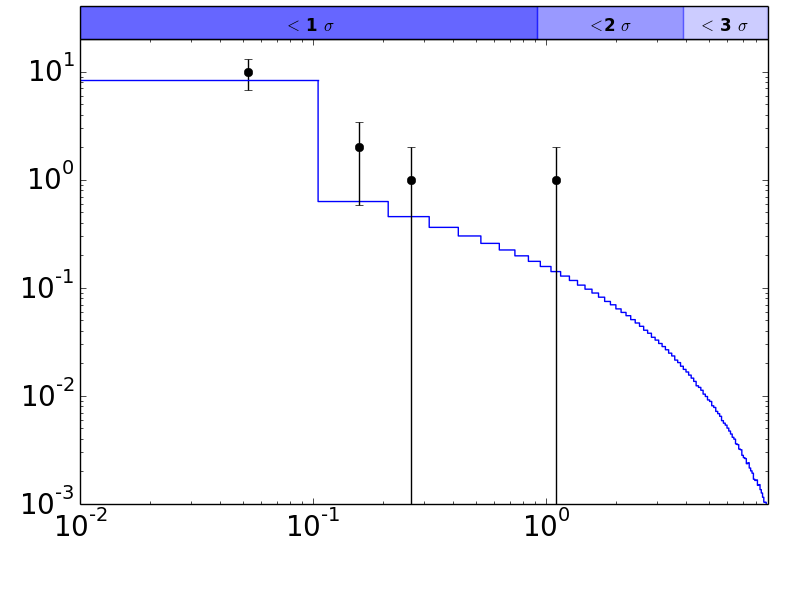}{0.33\textwidth}{(c)}
          }
\caption {The binned $\Lambda$ values obtained for each of the 14 (13 for UPMUs)
  search directions for (a) FC, (b) PC, and (c) UPMU topologies. Also
  shown is the expected test statistic distribution as predicted from
  our MC simulation. The confidence levels are determined by
  enclosing 68.3 \%, 95.4 \%, and 99.7\% of the test statistic
  prediction for 1 $\sigma$ (dark blue), 2 $\sigma$ (medium blue), and
  3 $\sigma$ (light blue), respectively. \label{fig:lambda_results}}
\end{figure*}

Figure \ref{fig:lambda_results} shows the
distribution of the test statistic separately for each of the three topologies
considered, along with the expected test statistic distribution
calculated using simulated data from our Monte Carlo code. As seen
here, no statistically significant excesses are seen and the
distribution of the test statistic from the data matches the
background expectation. The confidence levels are determined using the
MC test statistic distribution by calculating the $\Lambda$ where
68.3\%, 95.4\%, and 99.7\% of the test statistic prediction is enclosed for 1 $\sigma$, 2 $\sigma$, and 3 $\sigma$, respectively.

\subsection{Searching for a coincidence with the IceCube multiplet
  event}\label{subsec:multiplet}

In addition to the search for excesses in the direction of the IceCube
track events, we separately searched for coincidence events with the
IceCube multiplet event reported in \citet{aartsen17}. On February 17,
2016, IceCube observed three neutrino candidate events within less
than 100s separated by 3.6$^{\circ}$. This type of multiplet event
would be expected to occur once every 13.7 years. No optical
counterparts were found in the follow up searches discussed in
\citet{aartsen17}.

In SK, we searched for neutrino candidate events in the FC, PC, and
UPMU data sets in a 500-s time window around the time of the first
detected neutrino in the multiplet event. No SK candidate events were
detected in any of the three topologies.

We also performed a spatial coincidence search over all SK phases
using the same method used for the IceCube track events reported in
the other sections of this paper. In the direction of the triplet
event (dec = 39.5$^{\circ}$, RA = 26.1$^{\circ}$), we detected
(expected) 24 (24.5) FC events, 26 (23.2) PC events, and 16 (12.9)
UPMU events. The likelihood ratio ($\Lambda$) for the three topologies
was calculated to be 0 for FC, 0.14 for PC, and 0.30 for UPMU, which
is all less than the one sigma value determined using the atmospheric
MC background. 

\section{Conclusions} \label{sec:conclusions}

We performed a search for SK neutrino detections coincident in
direction with the IceCube track events. Using Poisson statistics, we
used SK data taken from April 1996 - April 2016 to
determine if there was any excess of events in each of the search
directions. The detected numbers of SK
neutrino events in each of the search directions were consistent with
the background expectations. The most significant $\Lambda$ was 1.1
at a declination of -22$^\circ$ in the UPMU data set, which
corresponds to a significance of about 1.1 $\sigma$ using the
atmospheric MC prediction.

We also looked for coincidence events with the IceCube multiplet event
reported in \citep{aartsen17}. In the time coincidence search, no
events were found within a $\pm$ 500 s time window from the first
detected IceCube event. In the direction coincidence search, the
number of events detected over the lifetime of SK from the direction
of the IceCube multiplet event was consistent with atmospheric
background for the FC, PC, and UPMU topologies.

These results represent the first search for coincident neutrinos with
IceCube events that has been extended down to the few GeV energy
regime. Given the unknown origin of these astrophysical neutrinos, it
is worth exploring all available data in the hopes of a new
discovery. This search was not optimized for a particular energy
spectrum, which trades improved sensitivity to the popular energy
spectra (E$^{-2}$, for example) for the flexibility of being model
independent. This search is useful for constraining the behaviour of
astrophysical neutrinos in the lower energy regime and guiding new
models which predict neutrino behaviours at lower energies.

\section{Acknowledgements} \label{sec:acknowledgements}

We gratefully acknowledge the cooperation of the Kamioka Mining and
Smelting Company. The Super‐Kamiokande experiment has been built and
operated from funding by the Japanese Ministry of Education, Culture,
Sports, Science and Technology, the U.S.  Department of Energy, and
the U.S. National Science Foundation. Some of us have been supported
by funds from the National Research Foundation of Korea
NRF‐2009‐0083526 (KNRC) funded by the Ministry of Science, ICT, and
Future Planning, the European Union H2020 RISE‐GA641540‐SKPLUS, the
Japan Society for the Promotion of Science, the National Natural
Science Foundation of China under Grants No. 11235006, the National
Science and Engineering Research Council (NSERC) of Canada, the Scinet
and Westgrid consortia of Compute Canada, and the National Science
Centre, Poland (2015/17/N/ST2/04064, 2015/18/E/ST2/00758).

\end{document}